\documentstyle[preprint,aps,tighten,prl]{revtex} 
\begin{document} 
\draft 
\preprint{CWRU-P12-95, CFPA-95-TH-26 } 
\title{Chaos, Fractals and Inflation} 
\author{ Neil J. Cornish$^{\dag}$ and Janna J. Levin$^{\star}$} 
\address{$^{\dag}$Department of Physics, Case Western Reserve University
\\ Cleveland, OH 44106-7079} 
\address{$^{\star}$Center for Particle Astrophysics,
UC Berkeley\\
 301 Le Conte Hall, Berkeley, CA 94720-7304} 
\maketitle 
\begin{abstract} 
In order to draw out the essential behavior of the universe, investigations
of early universe cosmology often reduce the complex system to a simple 
integrable system. Inflationary models are of this kind as they
focus on simple scalar field scenarios with correspondingly simple dynamics.
However, we can be assured that the universe is crowded with many interacting
fields of which the inflaton is but one. As we describe,
the nonlinear nature of these interactions
can result in a complex, chaotic evolution of the universe. Here
we illustrate how chaotic effects can arise even in basic models such as 
homogeneous, isotropic universes with two scalar fields. We find
inflating universes which act as attractors in the space of initial
conditions. These universes display chaotic transients in their early
evolution. The chaotic character is reflected by the
fractal border to the basin of attraction. 
The broader implications are likely to be felt in the process
of reheating as well as in the nature of the cosmic background radiation.

\end{abstract} 
\narrowtext 

\section{Introduction}

The inflationary paradigm strives to deliver a smooth universe
from random initial conditions. If inflation is a robust attractor
in the space of initial conditions, then it earns its
claim to naturalness and genericity \cite{{bel},{gp}}. 
A universe which hosts many different fields,
including an inflaton candidate, can develop an extreme
sensitivity to initial values. This 
sensitivity marks the onset of chaos.
Chaotic dynamics does not in itself destroy the robustness of an 
inflationary phase. Rather, it can lead to some powerful
and perhaps observable implications for a realistic universe.
For instance, a fractal pattern in the spectrum of density
fluctuations could be generated. Also, the final phase of
inflation marked by reheating would unavoidably
be a setting for chaos.

In simple cosmologies, the ultimate 
fate of the universe can be predicted once a set of initial
conditions is prescribed. In a closed cosmology for instance, it can be
determined from the initial prescription if the universe
inflates or collapses. A plot in phase space will show
regions or basins within which all of the initial
conditions lead to the same outcome. There will be 
basins of inflation and basins of collapse. If the 
dynamics is {\it not} chaotic, these basins of attraction\footnote{
Throughout the paper we loosely refer to attractors in phase
space. In more formal terminology, the word attractor is 
reserved for dissipative systems. Strictly speaking, our attractors
are just asymptotic regions of phase space.}
are distinctly separated by smooth, regular boundaries.
If the dynamics becomes chaotic, then the smooth 
boundaries begin to break up, ultimately becoming 
fractal.

The models described in the following sections
are chosen on the grounds of simplicity, and do not
necessarily conform to any standard inflationary scenario. In our
current models the primary inflaton is weakly coupled
and essentially not dynamical, leaving the chaotic dynamics to the
other scalar fields. Multiply coupled fields are quite
natural in any particle theory. The Higgs field for instance
must couple to the standard model fermions in order to induce fermion masses.
In supersymmetric theories, a glut of coupled particles is 
expected. The fields behave effectively like nonlinearly coupled 
harmonic oscillators and so naturally bring on chaos.
In future studies we intend to connect more concretely to specific
hybrid inflationary models\cite{alin} and draw out
implications for the spectrum of fluctuations or the end of a realistic model.
The main aim of this paper is to illustrate the complex dynamics that can
arise in relatively simple cosmologies.

In addition to highlighting
the appearance of chaos in inflationary cosmology, we aim 
to demonstrate the power of fractals as a quantitative 
measure of chaos in relativity.
In general relativity coordinate independent
measures of chaos are of vital importance. One of 
the most valuable measures of a chaotic system
in flat space, the Lyapunov exponents, can be removed by
a simple coordinate
transformation in curved space. Thus the usual coordinate dependent 
measures of chaos become ambiguous in a relativistic context.
Topological signals such as fractal basins, cantori or
stochastic layers in phase space are needed for conclusive evidence of 
chaos. In order to search for cantori or stochastic layers
it is necessary to construct slices through phase space known as Poincar\'{e}
sections. Since Poincar\'{e} sections rely on quasi-periodic behaviour, the
system must complete many cycles for a useful picture to emerge.
Oftentimes relativistic systems are not so obliging as the evolution 
may end at singularities, such as the big crunch or inside
a black hole. In these cases the dynamics is better suited to an outcomes based
approach such as the study of fractal basin boundaries\cite{dfc}.

In a chaotic system the different possible outcomes will
each have a basin of attraction in the space of initial conditions, with
the basins separated by a fractal border. 
A specific examination of phase space will require a coordinate
system to be chosen. One might worry that in a different time slicing,
the fractal would disappear. This is not possible.
A coordinate transformation must be smooth and differentiable.
No smooth map can undo a truly fractal pattern as fractals are
non-differentiable. While the
features of a fractal may be altered by a 
coordinate transformation, the existence of a the fractal
is unambiguous.

The cosmological context we employ allows us to demonstrate
the technique of fractal basin boundaries. What emerges is
a definitive manifestation of chaos in cosmology.
We have already commented that we focus on multi-field systems.
If only one scalar field is present and the universe is closed, 
then there is still the possibility of chaotic regions in phase
space \cite{{don},{cal1},{cal2}}. However, in order to 
generate chaotic dynamics with just one matter field, the universe
must oscillate between expansion and contraction many times.
The requirement of many bounces makes these otherwise interesting
solutions unlikely
if not truly unphysical. We consider single field scenarios
in \S  \ref{single}.  
Since these bouncing candidates do not represent
viable cosmologies, we turn our attention to many field systems
and the demonstration of fractal
basin boundaries in \S \ref{two}-\ref{diss}.  
Also note, that while we study closed
cosmologies, the chaotic transients can be seen in a universe
which never collapses or bounces. When fields interact,
the chaotic nature is therefore not limited to closed cosmologies.
It is thus possible that there was
a transient chaotic epoch in the history 
of our own universe.

\section{The Cosmological Model}

In the coming examples we consider closed Friedman-Robertson-Walker
(FRW) universes. For potential
driven inflation to be successful the inflationary potential needs to be
fairly constant. For our purposes the inflaton can be modeled
by a simple cosmological constant.
If this were the complete system there would be of course
no chaotic dynamics. However, the universe is created
bursting with matter fields. 
We model the matter content by a variety of conformally
and minimally coupled fields. These matter fields interact 
and can incite chaos. For the inflationary cases at hand then,
the chaotic behaviour is principally matter driven,
i.e. chaos in $T_{\mu\nu}$ causing chaotic evolution
of $g_{\mu\nu}$.

The physical picture is that of a closed, pre-inflationary universe just
exiting the Planck era.
The space of initial conditions is probed by assigning
three possible outcomes. Either the 
universe inflates forever, inflates for short spurts
but then collapses, or collapses without any inflationary
event. The outcome depends on the relative
sizes of the various kinetic and potential energies in the matter fields. 
As the interaction between the matter fields is turned
up the boundary which separates inflating from noninflating initial 
conditions blurs, eventually becoming fractal.

While we only consider closed models in our outcomes based
approach, the appearance of chaotic transients will be generic,
regardless of curvature.  This becomes clear since 
chaos is also nested within a given outcome basin. 
For instance, a universe will often go through rocky
beginnings, enduring many fits of inflation before taking
off smoothly or collapsing. Within a collapse basin,
the sensitivity to the initial conditions shows up as 
a random scatter in the maximum radius of the universe 
or in the final value of the fields.
If the universe has managed
to inflate by several e-folds 
there is no turning back as the kinetic energy
which might interrupt inflation quickly redshifts away. 
This feature of the
de Sitter attractor is often referred to as cosmic baldness. We shall see that
while the de Sitter attractor might end up bald, it can have very hairy
beginnings.

We shall consider FRW universes described by the
metric
\begin{eqnarray}
ds^2&=&dt^2-a^2\left({dr^2 \over 1-k r^2}-r^2 d\Omega^2\right) \; ,
\nonumber \\
&=&a^2\left(d\tau^2-{dr^2 \over 1-k r^2}-r^2 d\Omega^2\right) \; , 
\end{eqnarray}
where $t$ is cosmic time and $\tau$ is conformal time.
Throughout, 
dots will denote derivatives with
respect to cosmic time and dashes will
denote derivatives with respect to
conformal time. 
The matter Lagrangian will contain various combinations of conformally,
$\Psi$, and minimally, $\Phi$, coupled scalar fields with a variety
of interaction terms described by the potential $V(\Psi,\Phi)$:
\begin{equation}
{\cal L}_M=V-\frac{1}{2}\partial_{\mu}\Phi\partial^{\mu}\Phi-\frac{1}{2}
\partial_{\mu}\Psi\partial^{\mu}\Psi+\frac{1}{12}{\cal R}\Psi^2 \; .
\label{lagm}
\end{equation}
For comparison with the conformally coupled term,
i.e. the last term in eqn (\ref{lagm}), the
gravitational Lagrangian is ${\cal L}_G=-{1\over 12}{\cal R}$.
We have chosen units where $4\pi G/3=c=1$. 
In terms of cosmic time $t$ the field equations read
\begin{eqnarray}
&&\ddot{\Phi}+3H\dot\Phi+\partial_{\Phi}V=0\; , \label{cos}
\\ \nonumber \\
&&\ddot{\Psi}+3H\dot\Psi+\partial_{\Psi}V
+{{\cal R}\over 6}\Psi=0\; ,
\\ \nonumber \\
&&\ddot{a}+a\left(2\dot{\Phi}^2+(\dot{\Psi}+H\Psi)^2
+{k \over a^2}\Psi^2+\Psi\partial_{\Psi}V-2V\right)=0 \; .
\label{accel}
\end{eqnarray}
The Ricci scalar can be related to the scale factor through
${\cal R}/6={\ddot{a}/ a}+H^2+{k / a^2}$.
The Hubble expansion factor is given by $H=\dot{a}/a$.
The total energy of the system is
\begin{equation}
{\cal H}=H^2+{k \over a^2}-\dot{\Phi}^2-(\dot{\Psi}+H\Psi)^2
-{k \over a^2}\Psi^2 -2V =0 \; .
\label{ham}
\end{equation}
The constraint eqn (\ref{ham}) can be obtained
directly from Einstein's field equations
and represents the first integral of eqn (\ref{accel}).

Since we are dealing with the entire universe, the system is necessarily 
conservative. We use eqn (\ref{ham}) to ensure that
energy is in fact conserved. It is amusing to notice that if 
we isolate the matter sector, this subsystem looks dissipative.
Energy is lost to the gravitational field through the friction terms
$\sim 3 H \dot \Phi$. We can see the effects of dissipation within 
the larger context of the Hamiltonian system. For instance
we can watch the matter trajectories shrink down onto 
an attractor as the volume in phase space is dissipated 
(cf \S \ref{diss}).

In some cases it is profitable to recast the field equations in terms of
conformal time $\tau$, and the rescaled variables $\psi=a\Psi$, $\phi=a\Phi$
and $U=a^4V$:
\begin{eqnarray}
&&\phi''-{a'' \over a}\phi+\partial_{\phi}U=0 \\ \nonumber \\
&&\psi''+k\psi+\partial_{\psi}U=0 \\ \nonumber \\
&&a''+ka+{1 \over a}\left(\phi'-{a' \over a}\phi\right)^2-
{1 \over a}\left(4U-\psi\partial_{\psi}U\right)=0 \; .
\label{eom}
\end{eqnarray}
The total energy of the system can be expressed as
\begin{equation}
a^4{\cal H}=(a')^2+ka^2-\left(\phi'-{a' \over a}\phi\right)^2
-\left((\psi')^2+k\psi^2\right)-2U =0 \; .
\end{equation}
A universe is said to inflate if the scale factor $a$ accelerates in terms
of cosmic time, i.e. $\ddot{a}>0$. The cosmic time acceleration is given by
\begin{equation}
\ddot{a}={a a'' -(a')^2 \over a^3}=
{1 \over a^3}\left[2U-\psi\partial_{\psi}U-(\psi')^2-k\psi^2
-2\left(\phi'-{a'\over a}\phi\right)^2\right] \; .
\end{equation}
Notice that conformally coupled fields tend not to contribute to a
positive acceleration. In other words, even with a potential
$\propto \Psi^n$, conformally
coupled fields do not drive inflation (unless $n<2$).
The generalization of the above equations to describe two or more scalar
fields of either type is direct.

The field equations are invariant under the combined rescaling
\begin{equation}
a\rightarrow x a,\quad t\rightarrow xt,\quad V\rightarrow {V \over x^2}, 
\end{equation}
where $x$ is a constant.
This freedom is removed when we set our length scale by choosing
dimensionfull values for quantities such as masses and coupling constants.

As we describe below, the asymptotic solutions are of two kinds.
The universe eventually approaches a smooth de Sitter phase or
it ultimately collapses into a big crunch.
In highly simplified models the division between these two outcomes
can be expressed as a simple partition in the space of initial conditions.
However, we shall see that even in simple models with two 
interacting fields the
division is no longer clean, as the boundaries separating
the different outcomes are no longer smooth curves but fractals.

\subsection{From integrability to chaos}\label{road}

When the various scalar fields are massless and do not interact the
equations of motion can be integrated exactly and there is no chaos.
Before launching into the chaotic dynamics, we take a look in this
subsection at the two asymptotic possibilities,
the big crunch and de Sitter expansion, which will be the
basis of our outcomes approach
in the following sections. We also show
the phase space portraits for the non-interacting, closed system.
In Ref. \cite{bel}, a detailed analysis was given of the phase
space portraits for a single, massive, minimally coupled field
in a universe with arbitrary curvature. According to their
portraits, the trajectories drawn are untangled and therefore are
not chaotic. As the authors noted, there do
exist a set of measure zero oscillatory solutions which do show chaotic
behavior \cite{don}. At the close of \S \ref{single} we return to
discuss this special set of perpetually bouncing solutions.

For now we demonstrate the non-chaotic phase space for a universe
full of noninteracting, garden variety fields
(both minimally and conformally coupled).
Since it is impossible to tell one scalar field from another in the absence
of interactions, the general case of $N$ minimally coupled and $M$
conformally coupled scalar fields reduces to a universe with one field of
each type. For a universe with a scalar field of
each type and cosmological constant $\Lambda$ 
the equations simplify to
\begin{eqnarray}
&& \psi= A\, \cos\, \tau \, ,\\ \nonumber \\
&& \left({\phi \over a}\right)^{'}= {B \over a^2 } \, ,\\ \nonumber \\
&& a^{'}= \pm \sqrt{A^2+2 a^4 \Lambda +{B^2 \over a^2}-a^2} \, .
\end{eqnarray}
The phase space is divided by a separatrix into two classes of
trajectories, those that terminate at the big crunch and those that
inflate. 
The form of the solutions can easily be found in the 
neighborhood of these two geometrically distinct attractors.
When the dynamics is dominated by
potential terms, such as a cosmological constant $\Lambda$, the universe
undergoes exponential expansion and matter fields are redshifted away:
\begin{equation}
a \sim e^{\sqrt{2\Lambda}\ t}\sim - \left({1\over \sqrt{2\Lambda}\ \tau}
\right)\; 
, \quad \phi
 \sim a^{-2} \; , \quad \psi\sim  \cos\, \tau \,  .
\end{equation}
Conversely, when the dynamics is dominated by the kinetic energy of the
matter fields or spatial curvature, the universe collapses to the big crunch
at time $\tau_c$:
\begin{equation}
a \sim |\tau-\tau_c|^{1/2}\; , \quad \phi \sim {a \ln a} \; ,
\quad \psi\sim \cos \, \tau
\end{equation}
for $B\ne 0$ and for $B=0$:
\begin{equation}
a \sim |\tau - \tau_c |\; , \quad \phi \sim {\rm constant} \; ,
\quad \psi\sim \cos \, \tau \; .
\end{equation} 
The separatrix that partitions these possibilities is defined by the
trajectory with
\begin{equation}
\Lambda={A^2+2\sqrt{A^4+3B^2} \over 6 (A^2+\sqrt{A^4+3B^2})^2 } \, .
\end{equation}
For these simple, integrable cosmologies the basins of attraction for
the big crunch and de Sitter attractors are separated by a smooth
curve. This smooth curve is a portion of the separatrix. In Fig. 1
we display phase space portraits in the $(a,a')$ and $(\psi,\psi')$
planes for a universe with $B=0$ and $\Lambda=1/8$. The cross-hatched region
is the basin of the big crunch attractor and the solid line
is the separatrix.

When interactions are included the separatrix breaks up and is replaced by
a fractal curve. The gaps in the broken separatrix have the structure of a
Cantor set. The broken separatrix no longer partitions phase space and
trajectories may diffuse through it. For example, a universe that
was destined to collapse in the integrable case might diffuse through the
broken barrier and inflate. The breaking of the separatrix is reflected in
the fractal nature of the basin boundaries for chaotic universes. The
smooth basin boundaries shown in Fig. 1 should be compared to the fractal
boundaries seen in Figs. 4 and 7. The break up of the separatrix is
further described in \S \ref{two}.

Even when interactions are included, the asymptotic behaviour of
trajectories on either attractor is completely regular and non-chaotic.
Examples of this fact are given in \S \ref{diss}.
In the parlance of dynamical systems theory, the attractors are
neither strange nor chaotic. The chaotic behaviour is transient\cite{ott}, and
occurs when trajectories approach the broken separatrix. Physically
this corresponds to an epoch in which the universe coasts with
a fairly constant radius but with wildly varying acceleration. During this
epoch the kinetic and potential energies in the system
fight for supremacy and the universe teeters between collapse and violent
expansion. If the kinetic energy wins the day, the universe
collapses and the asymptotic solution can be found by
neglecting all potential terms in the equations of motion. Conversely,
if the potential energy wins the day the asymptotic solution can be found by
neglecting kinetic energy terms.

The transient nature of the chaos is similar to that found in the
Mixmaster universe\cite{{mix},{barrow}}, 
where it has been shown that the underlying
attractors are neither strange nor chaotic\cite{not}. We remark that
transient chaos appears to be the hallmark of relativistic systems.

\section{Cosmologies with a single scalar field}\label{single}

Since we work with a closed
FRW cosmology, there is only one parameter
describing the gravitational sector -- namely, the scale 
factor. The necessary elements for chaos are present
if the scale factor interacts even with just
one matter field.
However, the dynamical timescale for the onset
of chaos is longer than
the life of one universe.
The chaotic dynamics results as the 
two oscillators interact. Typically, at least a few
oscillations are needed for the effects
to surface. We discuss such an example in this section. 
On the other hand,
if there are many interacting matter fields in the universe,
then their chaotic evolution will make an impact
during the lifetime of one universe.
The examples of the following section reveal chaos on such 
short time scales.

We begin with an example that is chaotic, but only on a timescale longer
than the life of one universe. The model describes
a single, conformally coupled scalar field in a closed $(k=1)$ universe. We
choose the potential to have both a mass term and a cosmological constant
$\Lambda$:
\begin{equation}
U=\frac{1}{2}m^2 a^2\psi^2+a^4\Lambda \; .
\end{equation}
This example has previously been considered by Calzetta and El Hasi
\cite{cal1,cal2}.
The Hamiltonian takes the form
\begin{equation}
a^4{\cal H}=(a')^2+a^2-((\psi')^2+\psi^2+m^2 a^2\psi^2)-2a^4\Lambda=0 
\; ,
\end{equation}
which, aside from the wrong sign for the gravitational contributions,
is the Hamiltonian for two coupled harmonic oscillators.

For a metric with only one dynamical degree of freedom it is always
possible to perform a combined field redefinition and conformal transformation
to a coordinate system in which the dynamics appears to be nonsingular.
By using conformal, rather than cosmic time to describe the evolution of
this system, the dynamical equations can be smoothly integrated past the
big bang and big crunch singularities at $a=0$. This allows many cosmic
cycles to be considered if we continue the scale factor
into negative values. When evolved through a series of cosmic cycles the
system is clearly chaotic\cite{cal1,cal2}, as we might expect for nonlinearly
coupled oscillators. It should be noted that the cosmic cycles are
physically meaningless as all memory of the previous cycle is erased at
each big crunch singularity.

By introducing the fiction of cosmic cycles, the dynamics can be surveyed
using the standard tools of Poincar\'{e} sections (return maps) and Lyapunov
exponents. Lyapunov exponents measure the rate of separation of trajectories
in phase space. Only if trajectories separate exponentially fast do they have
positive exponents. Systems with positive Lyapunov exponents are said to
exhibit sensitive dependence on initial conditions - one of the two ingredients
of chaos (the other being the mixing and folding of trajectories). The inverse
of the positive Lyapunov exponents is referred to as the Lyapunov timescale.
This timescale sets the dynamical timescale over which chaotic effects make
themselves felt. In general relativity, Lyapunov exponents must be used with
extreme care, if at all, as they are coordinate dependent. Indeed, a simple
coordinate transformation can give a non-chaotic system positive exponents
and a chaotic system vanishing exponents.

Putting these reservations aside, we may compare the Lyapunov time to the
time taken to complete a cosmic cycle, and infer whether or not chaotic
effects can
make themselves felt in the lifetime of a single universe. Typically, the
Lyapunov timescale was found to be in the range $10\rightarrow 1000$ cosmic
cycles. Even when the mass is taken to be very large, the Lyapunov timescale
is always found to be greater than half a cosmic cycle, or in other words, the
timescale for chaos to become important always exceeds the life of one
universe. This result is easily understood. The chaotic behaviour is
due to resonances between the two oscillating fields $a$ and $\psi$. 
In order for the resonance to take effect, both fields typically need to
oscillate several times. However, $a$ can only complete half an oscillation
before the big crunch, making it exceedingly difficult for a chaotic resonance
to occur.

In Ref.\cite{cal1} it was argued that chaos had been viewed within
the span of one life cycle. Their conclusion was based on what
appeared to be a scatter between initial values of the matter fields
and the final values.
The correlation between initial and final values of the scalar
field $\psi$ was found to be $0.01$. 
However, this low value for the
correlation actually stems from a coarse sampling of a high frequency
function. By regenerating Fig. 6. of Ref.\cite{cal1} with
a sampling rate that is ten times higher we see from Fig. 2. that the true
correlation coefficient is $1.00$. This confirms that the
system shows no meaningful chaotic effects in the life of one universe.

Similar conclusions hold for universes inhabited by a single minimally
coupled scalar field. Again, the equations of motion can lead to
chaotic behaviour as they are nonlinear and have phase space dimension
greater than 2. However, meaningful chaotic effects can only occur if
the universe itself oscillates. The dynamics of an inflationary model
driven by a minimally coupled scalar field with potential
\begin{equation}
V=\frac{1}{2}m^2\Phi^2+{\lambda \over 4}\left(\Phi^2-\Phi_{0}^2\right)^2 \, ,
\end{equation}
was studied by Belinskii {\it et al.}\cite{bel}, and with $\lambda=0$
by Hawking\cite{hawk} and Page\cite{don}. Typical trajectories
were not chaotic. Rather, they see the universe smoothly evolve
from the big bang to the big crunch
with various amounts of inflation \cite{bel}. 
However, the inflationary potential
allows some atypical trajectories for which  
the universe undergoes a number of
nonsingular bounces \cite{hawk}. Page\cite{don} suggested that
there exist an uncountably infinite but discrete set of perpetually
bouncing universes with vanishing Lebesgue measure but non-vanishing
fractal dimension. If Page's suggestion is correct, it would prove
that the dynamics is chaotic as his ``fractal set of perpetually
bouncing universes'' corresponds to what is now known as a strange
repeller\cite{kad}. In contrast to the fictional cosmic cycles used to
describe a conformally coupled scalar field, Page's bouncing universes
are true, nonsingular solutions. However, these solutions have 
obvious drawbacks as plausible cosmologies.  
As remarked in
Ref.\cite{bel}, the fine tuning required to arrive at these chaotic
trajectories rules them out as a robust physical model displaying
chaotic behaviour. Perhaps in a model of the early universe that generically
displays nonsingular bounces we can hope to see interesting chaotic
effects caused by an oscillating scale factor. In the absence of such
a model we have to look to additional matter fields to provide the
nonlinear resonances needed to incite chaos.

\section{Cosmologies with two conformally coupled fields}\label{two}

If additional fields occupy the universe, then the scale factor
will not be the principle source of chaos.
Two scalar fields can oscillate many times in the lifetime of one
universe, leading to truly chaotic behaviour. 
To demonstrate the chaos we
show the fractal basin boundaries for a universe
which contains two conformally coupled fields which interact
through the potential
\begin{equation}
U=\frac{1}{2}m_{1}^2 a^2\psi_{1}^2+\frac{1}{2}m_{2}^2 a^2\psi_{2}^2
+\lambda^2\psi_{1}^2\psi_{2}^2+a^4\Lambda \; .
\end{equation}
The period of oscillation for each field is governed by its effective mass.
We define the reduced effective mass for each field as
the derivative with respect to the field of the field eqn (\ref{eom}).
In other words, 
$M^2_\psi$ has the form of $\partial^2 W/\partial \psi^2$
were $W$ is anything which acts as a potential in the equations 
of motion:
\begin{eqnarray}
&& M_{a}=(1-m_{1}^2\psi_{1}^{2}-m_{2}^2\psi_{2}^{2}-4\Lambda a^2)^{1/2} \; ,
\\ \nonumber \\
&& M_{1}=(1+m_{1}^2 a^2+2\lambda^2\psi_{2}^2)^{1/2} \; , \\ \nonumber \\
&& M_{2}=(1+m_{2}^2 a^2+2\lambda^2\psi_{1}^2)^{1/2} \; .
\end{eqnarray}
Increasing $m_{1},\, m_{2},\, \Lambda$ and $\lambda$ slows the recollapse
of $a$ and speeds the oscillation of $\psi_{1}$ and $\psi_{2}$, thus
increasing the probability of chaotic resonances. However, if $m_{1}$ or
$m_{2}$ greatly exceed $\lambda$, the resonances will be washed out and no
chaos will be seen. Conversely, if $m_{1}$ or $m_2$ are both zero, the
oscillations tend to freeze when $\psi_1$ and $\psi_2$ hit small values,
again making chaotic resonances unlikely.

To gain some intuition we can find a simple analytic 
approximation which corresponds to a familiar chaotic system.
During the majority of the universe's evolution, the scale factor
varies much more slowly than the scalar fields so that $a'/a \ll \psi_{i}'/
\psi_{i}$. The scalar fields behave like coupled nonlinear oscillators,
adiabatically pumped by the slowly varying scale factor. To leading
order we can ignore the adiabatic pumping all together and study the
scalar field dynamics in a fixed background $(a' \approx 0)$. This
approximation is particularly good for describing universes that are
vacillating between collapse and inflationary expansion. Importantly,
this is just the region where the chaotic transients occur that
destroy the smooth separatrix of the
integrable model described in \S \ref{road}. When
$a'\approx 0$ the dynamics simplifies to that of two coupled oscillators:
\begin{eqnarray}
&&\psi_{1}''+\omega_{1}^2 \psi_{1}+2\lambda^2\psi_{2}^2\psi_{1}=0 \, , \\
\nonumber \\
&&\psi_{2}''+\omega_{2}^2 \psi_{2}+2\lambda^2\psi_{1}^2\psi_{2}=0 \, ,
\end{eqnarray}
where $\omega_{i}^{2}=1+m_{i}^2 a^2$ is the fixed frequency of the
uncoupled $(\lambda=0)$ oscillators. The above system of equations
describes a known chaotic system\cite{quartic}, and the transition
to chaos as $\lambda$ is increased can be studied using the
Chirikov resonance overlap condition\cite{chirk}.
Having established that the fast variables $\psi_{1}$ and $\psi_{2}$
behave chaotically, we can then consider how they backreact on the
slow variable $a$. When looked at on timescales long compared to
the periods of the scalar fields, the evolution of the scale factor
is similar to Brownian motion, and can be described in terms of
chaotic diffusion equations\cite{ott}. It is this buffeting of the
scale factor by the matter fields that breaks the separatrix in
the $(a,a')$ plane and causes the universe to evolve in a chaotic manner.

Returning to the full, unapproximated equations we numerically
investigate the phase space of initial conditions.
For a given set of initial conditions we can identify three main
outcomes. The first possibility sees the universe expand and
collapse without 
any inflationary burst. The second
possibility sees the universe undergo one or many short bursts of
inflation, but failing to become a macroscopic universe. The third
possibility sees the universe sustain a prolonged and violent period
of inflation resulting in the formation of a macroscopic universe.
The first and second possibilities (coloured Black and Grey respectively)
combine to form the big crunch basin of attraction. This artificial
division of the big crunch basin is mostly for visual effect.
There is a fourth possible outcome that should be mentioned. There
are a set of trajectories with zero Lebesgue measure that
oscillate eternally, never  entirely collapsing
or reaching the de Sitter attractor.
These trajectories form the border between the big crunch and
de Sitter basins of attraction. We will see that these trajectories
belong to a fractal set of perpetually bouncing universes. In the
parlance of dynamical systems theory, this set forms the stable
manifold of a strange repeller\cite{ott}. 

The three possibilities are displayed graphically in Fig. 3 for the choice
of parameters $(\Lambda=0.0001,m_{1}=0,m_{2}=0.05,\lambda=1)$ and initial
conditions $\{a(0)=2,\, \psi_{1}=0.4,\, \psi_{2}=6,\, \psi^\prime_{2}=20\}$.
The initial values of $\psi_{1}'$ are $\{-23.31,\, -23.32,\, -23.33\}$, and
$a'(0)$ is fixed by the Hamiltonian constraint.

The fact that minute changes in the initial conditions can lead to such
dramatic changes in the outcome suggest that the fate of our model universe
is indeed chaotic. This suspicion can be confirmed by studying the boundary
between the basins of attraction of the three outcomes. Since the
basins are embedded in a six dimensional phase space, we are forced to
consider lower dimensional slices through the boundary. In Fig. 4 we display
a two-dimensional slice in the $\psi_{1}$ -- $\psi_{1}'$ plane for
universes with parameters and conditions identical to those used in Fig. 3.
The three basins of attraction (Black, Grey, White) are dramatically intermixed
{\em strange basins}, as at least a portion of the boundaries
are fractal. The boundaries near the origin are regular and smooth while
the outer boundaries appear fragmented. A detail of the outer region is
shown in Fig. 5, visually confirming the fractal nature of the boundary.
Repeated magnification reveals similar striated pictures on all scales.

Rather than rely on these qualitative features, we may
quantify the fractal nature of the boundary in terms of the fractal dimension.
There are many definitions of fractal dimension that we may choose from,
but the one best suited to our situation is the box counting dimension. On
a two-dimensional slice through phase space we cover the fractal with a grid
of squares with side length $\varepsilon$. We then count the number,
$N(\varepsilon)$, of squares needed to cover the fractal, {\it i.e.} the
number of squares containing more than one colour. The box dimension $d_{B}$
is defined by
\begin{equation}
d_{B}=-\lim_{\varepsilon \rightarrow 0}{\ln N \over \ln \varepsilon} \; .
\end{equation}
For self-similar structures the formal limit $\varepsilon\rightarrow 0$ need
not be taken, and in all practical situations we are only interested in the
existence of such scaling laws over a large, but not necessarily infinite,
range of scales. Since the fractal dimension is not invariant under
homeomorphisms, it is not a true topological invariant. However, it is
invariant under diffeomorphisms, so it does provide a topological measure in
general relativity. {\em The existence of fractal structures in
phase space provides a coordinate independent signal of chaos in relativity}.

The importance of the fractal dimension of the basin boundaries can be
described in terms of final state sensitivity\cite{gre}. Consider an initial
configuration near the basin boundary, where the uncertainty in the initial
conditions describes an N-dimensional ball of radius $\delta$ in the
N-dimensional phase space. The final state sensitivity $f_{\delta}$ is
the fraction of phase space volume which has an uncertain outcome due to
the uncertainty in the initial conditions, and is given by
\begin{equation}
f_{\delta}=\delta^{\alpha}\; ,\quad  \alpha=N-d_{B}  \; .
\end{equation}
For a non-chaotic system $\alpha=1$ and the final state sensitivity is directly
proportional to the initial uncertainty. For chaotic systems however,
$0<\alpha<1$, and the uncertainty in the outcome is greater than the
uncertainty in the initial conditions. For example, if $\alpha=0.47$, a
$50\%$ reduction in the initial uncertainty only reduces the final state
uncertainty by $28\%$. In this way, the dimension of the basin boundary
is a direct measure of ``sensitive dependence on initial conditions''.

In Fig. 6 we display the plot used to determine the fractal dimension of
Fig. 5. Because the three boundaries are densely interwoven in this case,
we chose only to calculate the dimension of the boundary between the big crunch
and de Sitter attractors, {\it i.e.} counting Grey and Black as one basin.
Using an $840\times 840$ grid we found the dimension to be $1.58\pm 0.02$.
The grid size of $840=2^3\times 3\times 5 \times 7$ was chosen as it has the
most factors of any number below $1000$. The curvature of the data points at
small and large $\varepsilon$ is to be expected.
For large $\varepsilon$ the covering is very inefficient, while for small
$\varepsilon$ the squares saturate the resolution used to generate the fractal.
These effects cause $d_{B}$ to tilt toward $2$ for large $\epsilon$ and
toward $1$ for small $\epsilon$.
Despite these limitations, accurate fractal dimensions can be obtained very
quickly and easily. For different choices of parameters we found fractal
dimensions ranging from $1$ to $1.96$, essentially filling the allowed range
$d_{B}=[1,2]$.

The boundary was found to be fractal on all possible two-dimensional slices.
For example, in Fig. 7, the boundary is shown in the $a$ -- $a'$ plane for a
slice which intersects Fig. 3 along the line $\psi_{1}=1.0$. The fractal
dimension of this slice was found to be $d_{B}=1.37\pm 0.02$. 

While the previous chaotic
pictures were typical of those found, the dynamics of
the system is not always chaotic. For small values of 
$\lambda,\, m_{1},\, m_{2}$ 
(at fixed scaling $x$) the dynamics is near integrable and the basins
are not strange, but regular. In Fig. 8 we increment $\lambda$ while keeping
all other parameters and initial conditions fixed. The mixing of the basins
is reminiscent of the blending of viscous fluids.
The dimension of the basin boundary for $\lambda=0.5$ was found to
be $d_{B}=0.99\pm 0.02$, which is consistent with a dimension of 1.
So, within errors, this boundary is smooth and non-chaotic.
To compare, the dimensions of the boundaries for $\lambda=2.0$
was $d_{B}=1.16\pm 0.05$ (Grey-White) and $d_{B}=1.26\pm 0.05$ (Grey-Black).

An important property of dynamical systems with strange attractor basins
is that the chaotic dynamics is not restricted to phase space trajectories
near the fractal boundaries. One way to see this might be to use the fiction
of cosmic cycles to follow the evolution of trajectories
starting in the big crunch basin. The Lyapunov
exponents and Poincar\'{e} sections for these trajectories would reveal
chaotic behaviour across the basin. However, we are not really interested
in effects which take longer than one universe's lifetime to make themselves
felt. Instead we plot in Fig. 9 the correlation between the initial
value of $\psi_{1i}$ and the value at the point of maximum expansion,
$\psi_{1m}$. The graphs are for a $\psi_{1}'=0$ slice through the big crunch
basin of Fig. 8 with $\lambda=2.0$. The big crunch basin stretches from
$\psi_{1}=0$ to $\psi_{1}\sim 2.9$ (and similarly for negative $\psi_{1}$).
A general increase in frequency with increasing $\psi_{1i}$ requires that we
use several plots, each covering half the region of the last, to cover the
basin. Unlike the regular plot seen in Fig. 2, the relationship
between initial and final values of $\psi_{1}$ is highly erratic, with
apparently random changes in frequency and amplitude.

\section{Cosmologies with minimally and conformally coupled fields}
\label{diss}

The chaotic behaviour seen in the previous system is not restricted to
conformally coupled fields. Similar behaviour is found for minimally coupled
fields with the same choice of potential. The main difference in this case
comes from the scalar fields themselves being a source of inflation,
in addition to the cosmological constant. The acceleration in this example is
given by
\begin{equation}
\ddot{a}=2a\left(\Lambda +\frac{1}{2}m_{1}^2\Phi_{1}^{2} 
+\frac{1}{2}m_{2}^2\Phi_{2}^2 +\lambda^2\Phi_{1}^2\Phi_{2}^2
-\dot{\Phi}_{1}^2-\dot{\Phi}_{2}^2\right) \; ,
\end{equation}
where we have reverted to the unscaled field variables. These models are able
to successfully inflate even when there is no cosmological constant, in
a manner similar to Linde's ``chaotic inflation''\cite{lin}. However,
we did not see any strange basins when $\Lambda=0$ as successful inflation
generally required the fields to become stuck high up in their potentials
after just a few oscillations. Otherwise, their ability to climb high enough
was lost due to friction and redshifting of kinetic energy. It may be that
chaotic behaviour does occur when $\Lambda=0$, but it is difficult to search
for as the inflationary bursts must be followed for $\sim 60$ e-folds in
comparison to the $\sim 5-10$ e-folds required to ensure we have
reached the de Sitter attractor when $\Lambda\neq 0$.

In Fig. 10 we display the basins of attraction in the $\Phi_{1}$ -
$\dot{\Phi}_{1}$ plane for universes with $(m_{1}=0,\, m_{2}=0.04,\,
\Lambda=0.00005,\, \lambda=1.0)$ and fixed initial conditions $\{a(0)=10.0,\,
\Phi_{2}=0.4,\, {\dot\Phi_{2}}=0.16\}$. A detail of the outer boundary is
shown in Fig. 11, where the dimension was found to be $1.54\pm 0.02$.
Because the scalar fields themselves contribute to the inflationary bursts
there is typically far more of the Grey basin than we saw for conformally
coupled fields.

We can compare an analysis of the chaotic trajectories with the 
non-chaotic trajectories of \S \ref{road}.
While the universe is expanding ($H>0$), we see from eqn (\ref{cos}) that
the scalar field dynamics is effectively that of a damped harmonic
oscillator. Conversely, as the universe contracts the dynamics is
that of a pumped harmonic oscillator. This behaviour is apparent in
Fig. 12, where we have displayed typical trajectories leading to the
de Sitter and big crunch attractors. For the de Sitter attractor the
scalar fields spiral into a fixed point as cosmic baldness asserts itself,
while for the big crunch attractor the scalar fields first spiral in
and then spiral out again as the universe collapses.
The de Sitter attractor has an interesting structure when two fields are
present as one scalar field gets locked at a constant value. The attractor
is of the form
\begin{eqnarray}
&&a=a_{c}\exp(\sqrt{2\Lambda}t) \; , \\ \nonumber \\
&&\Phi_{2}=\Phi_{2c}\exp\left(-{3\sqrt{2\Lambda} \over 2}t\right)
\cos\omega t \; , \\ \nonumber \\
&&\Phi_{1}=\Phi_{1c}\left[ 1 + \lambda^2\Phi_{2c}^2\exp(-3\sqrt{2\Lambda}t)
\left( {t \over 3\sqrt{2\Lambda}} + {2\omega\cos^{2}\omega t +3\sqrt{2\Lambda}
\cos\omega t \sin\omega t \over 2\omega (9\Lambda+2\omega^2) }\right)\right]
 \; ,
 \end{eqnarray}
 subject to the restriction
 \begin{equation}
 \omega^2=m_{2}^2+2\lambda^2\Phi_{1c}^2-\frac{9}{2}\Lambda \; .
 \end{equation}
For reference, the trajectory in Fig. 12a has $\Phi_{1c}=0.00573$, $\Phi_{2c}=
105.9$ and $\omega = 0.03796$. The big crunch attractor is unchanged from
the one-field case, and takes the form
\begin{eqnarray}
&& a=a_{c}(t_{c}-t)^{1/3} \, , \\ \nonumber \\
&& \Phi_{1}\sim \Phi_{1c}\ln(t_{c}-t) \, ,\\ \nonumber \\
&& \Phi_{2}\sim \Phi_{2c}\ln(t_{c}-t) \, ,
\end{eqnarray}
with $\Phi_{1c}^2+\Phi_{2c}^2=1/9$.
The trajectory shown in Fig. 12b has $t_{c}=200.004$, $a_{c}=13.68$,
$\Phi_{1c}=0.294$ and $\Phi_{2c}=0.167$. Since we are able to
write down analytic solutions for trajectories on the attractors, it is
clear that the de Sitter and big crunch attractors are non-chaotic. The
chaotic behaviour seen during the evolution of the universe
is restricted to the $5\rightarrow 10$ transient orbits seen in Fig. 12.
The fractal nature of the attractor
basin boundaries is due entirely to these brief chaotic transients.

We close with a word on a mixed
cosmology which  
contains one minimally coupled and one conformally coupled
scalar field. The conformally coupled scalar field $\Psi$ is taken to be
massless. When $\Psi$'s coupling to the minimally coupled scalar field
$\Phi$ is small it behaves like radiation. The interaction potential is
taken to be
\begin{equation}
V=\frac{1}{2}m^2\Phi^2+\lambda^2\Psi^2\Phi^2+\Lambda \; .
\end{equation}
As before, the minimally coupled scalar field is able
to contribute a negative pressure to that of the inflaton, thereby
increasing the likelihood of inflationary bursts. Again, this increases the
proportion of Grey over what we saw for two conformally coupled fields.

By viewing the basins in the $a$ - $\dot{a}$ plane we see some rather striking
ink-blot and crystal boundaries. An example of this is shown in
Fig. 13 for the choice of parameters $(m=0.05,\, \Lambda=0.0001,\,
\lambda=2)$,
and fixed initial conditions $\{\Phi=0.2,\, \Psi=0.1,\, \dot{\Psi}=0.08\}$.
A detail of the Grey-White crystal boundary is shown in Fig. 14. The
high degree of self-similarity of this fractal allowed a particularly accurate
determination of the fractal dimension using a standard $840\times 840$
grid. The dimension was found to be $d_{B}=1.484 \pm 0.005$. The visually
less fractal Grey-Black ink-blot boundary was found to have a
smaller fractal dimension of $d_{B}=1.11 \pm 0.05$, with the larger error
due to a lower degree of self-similarity.

In addition to studying different combinations of minimally and conformally
coupled scalar fields, we also considered a variety of polynomial
potentials $V(\Phi,\Psi)$. The qualitative results were the same for
all cases, showing that chaotic evolution was a generic feature of
all multi-field models.

\section{discussion}

The early universe is likely to be home to many interacting fields.
We have show that if these interactions are sufficiently strong, the
evolution of the universe will be chaotic. 
The fractal basin boundaries reveal the chaos in
a coordinate independent manner. Additionally,
the method does not require one universe to pass through
many cycles. We now have to ask how
prevalent chaotic behaviour will be in particular models of inflation,
and what implications it might have for processes such as reheating or
galaxy formation.
 
One model of inflation where chaotic dynamics is bound to be important
is hybrid inflation. Hybrid models employ several interacting
scalar fields, and arise naturally in various supersymmetric theories
where the breaking of large gauge groups employs many Higgs particles
\cite{randall}. If in the early stages the universe inflates in jolts, 
an exciting possibility exists for the spectrum of primordial density
fluctuations. Chaotic resonances could lead to a fractal power distribution,
perhaps helping to explain the hierarchical clustering seen in the current
universe. Moreover, chaotic evolution of the scale factor would leave
a unique imprint on the gravitational waves produced during inflation\cite{gr}.
Any chaotic behaviour would have to occur within the last 
$\sim 60$ e-folds of inflation to be observable today. This would require some
artificial fine tuning in a single field model but may be more natural in a
hybrid model.

Even in inflationary models where chaotic evolution is unimportant
at early stages, chaos is sure to play an important role 
at the end stages. To illustrate, consider again a hybrid model.
The fields to which the inflaton couples dictate the 
occurrence of the true vacuum and so control the nature of the
exit from inflation. The setting is prime for chaotic interactions which 
would certainly impact on the exit style. More generically, at the
end of any inflation model,
the universe reheats as the inflaton oscillates about the
minimum of its potential.
Particles are thereby produced through the inflaton's coupling to other
matter fields. If the matter fields are dynamical and chaos reigns, then the
process of entropy production would deserve rethinking.
The importance of parametric resonances, which are closely related to
chaotic behaviour, has already been stressed in this context\cite{{ll},{str}}.

It seems appropriate to consider how chaotic dynamics might impact
on ``chaotic'' inflation\cite{lin}. In chaotic inflation different patches of
the universe are taken to have different values of the inflaton and matter
fields. In some patch, it is argued, the inflaton is sufficiently
high up in the potential and the matter fields are sufficiently small 
so as to permit a long-lived inflationary epoch \cite{gp}. 
For initial field values
deep within a basin, away from the fractal borders, the usual arguments
hold and the chaotic inflation paradigm is largely unaffected.
However, if in a given patch the field values are
near a fractal basin boundary, it can become difficult to find a 
patch of any size across which the conditions are regular
enough to allow this thinking. Even the slightest variation in the
initial conditions across the patch will lead to an entirely different
outcome. Due to the self-similar nature of the fractal, no matter how small
you try to make the patch, there will still be slight variations in the
conditions and hence the outcome. For cosmological conditions in the vicinity
of the fractal basin boundary then, the simple FRW thinking must be abandoned. 

For similar reasons, caution would be needed for slow-roll 
initial conditions as well.  In fact the slow-roll scenario 
will likely be more fragile as the inflaton is more easily
kicked around.  For chaotic initial conditions by contrast the 
field high up in its potential is resiliant against
the influence of kicks and bumps.

Aside from direct observational effects there are also some important
theoretical implications raised by chaotic evolution. For example,
chaotic systems are characterised by an entropy,
the Kolmogorov-Sinai entropy, which is related to the spectrum
of Lyapunov exponents. This introduces a chaotic arrow of time in
addition to the cosmological and thermodynamic arrows of time. 
As well, it raises the question of a possible 
connection between the Kolmogorov-Sinai entropy and the thermodynamic
entropy released at the end of inflation.
Another issue raised by chaotic dynamics concerns the recovery of a
semiclassical limit in quantum cosmology due to the breakdown of the WKB
approximation in chaotic systems\cite{berry}.

We have suggested a few implications for chaotic dynamics in largely
unexplored terrain. Chaos theory grew out of Poincar\'{e}'s study
of the solar system, and with its development came insights into the 
intricate structures of our neighborhood such as the asteroid belt and
saturn's rings. It seems fitting for chaos to have an impact not only on
the evolution of the solar system but also on the birth of the universe.
We are left to ask if chaotic fingerprints have been left on the large-scale
landscape as they were on the landscape of our own solar system.

\section*{Acknowledgements} 
We have enjoyed discussions with Matt Holman and Jihad Touma. We thank
John Barrow, Norm Frankel, Andrei Linde and Don Page for their helpful
comments. We are indebted to Carl Dettmann for making his original programs
available to us, and to Jacques Legare for his help in making modifications.

\
\begin{figure}[h]
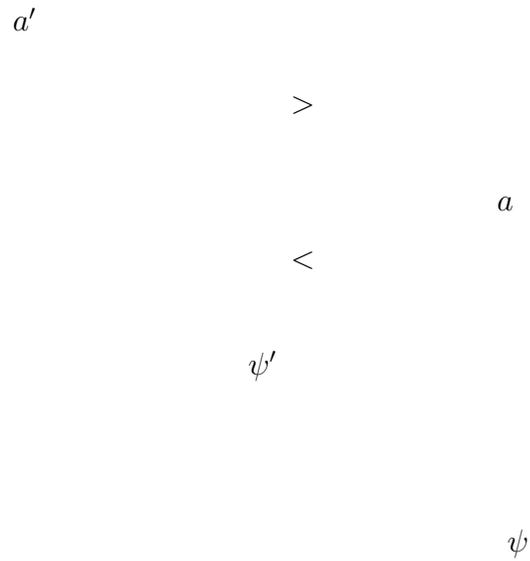

\vspace{70mm}
\includegraphics{blank.ps}
\includegraphics{phase1.ps}
\includegraphics{phase2.ps}
\vspace{30mm}
\caption{ Phase space trajectories in the $(a,a')$ and $(\psi,\psi')$ planes
for universes with non-interacting scalar fields. The solid line
is the separatrix and the cross hatched regions mark
the big crunch basin of attraction.}
\end{figure} 

\vspace*{-55mm}
\begin{picture}(0,0)
\put(139,190){$a '$}
\put(322,122){$a$}
\put(228,60){$\psi '$}
\put(326,-7){$\psi$}
\put(244,159){$>$}
\put(244,100){$<$}
\end{picture}

\newpage

\
\begin{figure}[h]
\vspace{70mm}
\includegraphics{blank.ps}
\includegraphics{calzetta3.ps}
\includegraphics{calzetta2.ps}
\vspace{20mm}
\caption{ The correlation between initial and final values of the scalar
field (a) is compared to the apparently chaotic behaviour seen in (b)
where the sampling rate is ten times lower.}
\end{figure}

\begin{figure}[h]
\vspace{70mm}
\includegraphics{ainflate4.ps}
\includegraphics{ainflate3.ps}
\includegraphics{ainflate5.ps}
\vspace{25mm}
\caption{The three possible outcomes for the universe. In each case
the solid line is the scale factor $a$ and the dashed line is the scaled
acceleration $\ddot{a} a^3$. The initial values for $\psi_{1}'$ are $-23.31$,
$-23.32$ and $-23.33$ respectively.}
\end{figure} 

\vspace*{-30mm}
\begin{picture}(0,0)
\put(58,455){$\psi_{f}$}
\put(214,326){$\psi_{i}$}
\end{picture}
\newpage

\
\begin{figure}[h]
\vspace{70mm}
\includegraphics{conformal1.out.ps}
\vspace{20mm}
\caption{ The basins of attraction for universes similar to those shown in
Fig. 3. }
\end{figure} 

\vspace*{5mm}

\begin{figure}[h]
\vspace{70mm}
\includegraphics{conformal2.out.ps}
\vspace{30mm}
\caption{ A detail of Fig. 4.}
\end{figure} 

\vspace*{-30mm}
\begin{picture}(0,0)
\put(38,445){$\psi_{1}'$}
\put(206,302){$\psi_{1}$}
\put(38,120){$\psi_{1}'$}
\put(206,-23){$\psi_{1}$}

\end{picture}
\newpage
\
\begin{figure}[h]
\vspace{60mm}
\includegraphics{conbox2.ps}
\vspace{10mm}
\caption{ Finding the fractal dimension for Fig. 5. The solid line is a
least-squares fit to the box counting data. The dimension was found to be
$d_{B}=1.58 \pm 0.02$.}
\end{figure} 

\vspace*{5mm}

\begin{figure}[h]
\vspace{70mm}
\includegraphics{adota.out.ps}
\vspace{35mm}
\caption{ A slice in the $a$ - $a'$ plane which intersects the $\psi_{1}$
- $\psi_{1}'$ plane of Fig. 4. along the line $\psi_{1}=1.0$.}
\end{figure} 

\vspace*{-30mm}
\begin{picture}(0,0)
\put(38,129){$a'$}
\put(206,-14){$a$}
\end{picture}

\newpage

\
\begin{figure}[h]
\vspace{70mm}
\includegraphics{/users/cornish/janna/ssp5.out.ps}
\includegraphics{/users/cornish/janna/ssp10.out.ps}
\includegraphics{/users/cornish/janna/ssp15.out.ps}
\includegraphics{/users/cornish/janna/ssp20.out.ps}
\vspace{20mm}
\caption{The road to chaos: As $\lambda$ is incremented from $0.5$ to
$2.0$, the nonlinear distortion of the attractor basin boundaries
mounts. Once $\lambda$ exceeds $1.0$ the mixing is so strong that the
boundaries become fractured, and eventually, fractal. The graphs were
generated for the choice of parameters and initial conditions
$\Lambda=0.00002,\, a_{0}=10,\, \psi_{2}=5.0,\, {\psi}_{2}'=10.0,\,
m_{1}=0.2,\, m_{2}=0.1$.}
\end{figure} 

\begin{figure}[h]
\vspace{70mm}
\includegraphics{/users/cornish/janna/psipsi1.ps}
\includegraphics{/users/cornish/janna/psipsi2.ps}
\includegraphics{/users/cornish/janna/psipsi3.ps}
\includegraphics{/users/cornish/janna/psipsi4.ps}
\vspace{20mm}
\caption{The correlation between $\psi_{1i}$ (vertical axis)
and $\psi_{1m}$ (horizontal axis) on a
$\psi_{1}'=0$ slice through the big crunch basin of attraction of Fig. 8.
$(\lambda=2.0)$.}
\end{figure} 

\vspace*{-30mm}
\begin{picture}(0,0)
\put(55,455){$\psi_{1}'$}
\put(210,312){$\psi_{1}$}
\end{picture}
\newpage

\
\begin{figure}[h]
\vspace{70mm}
\includegraphics{mini2.out.ps}
\vspace{20mm}
\caption{ Basins of attraction in the $\Phi_{1}$ - $\dot{\Phi}_{1}$
plane for universes containing two minimally coupled scalar fields. }
\end{figure} 

\vspace*{5mm}

\begin{figure}[h]
\vspace{70mm}
\includegraphics{mini3.out.ps}
\vspace{30mm}
\caption{ A detail of Fig. 10. where the dimension is $d_{B}=1.54\pm 0.02$.}
\end{figure} 

\vspace*{-30mm}
\begin{picture}(0,0)
\put(38,455){$\dot{\Phi}_{1}$}
\put(206,312){$\Phi_{1}$}
\put(38,120){$\dot{\Phi}_{1}$}
\put(206,-23){$\Phi_{1}$}
\end{picture}
\newpage

\
\begin{figure}[h]
\vspace{50mm}
\includegraphics{desitter.ps}
\includegraphics{bcrunch.ps}
\vspace{25mm}
\caption{(a) A trajectory spiraling into the de Sitter attractor. (b)
A nearby trajectory which flows out to the big crunch attractor.}
\end{figure} 

\vspace*{10mm}

\begin{figure}[h]
\vspace{70mm}
\includegraphics{mixprl1.out.ps}
\vspace{30mm}
\caption{ The basins of attraction in the $a$ - $\dot{a}$ plane
for universes containing a minimally coupled scalar field and a massless,
conformally coupled scalar field. }
\end{figure} 

\vspace*{-30mm}
\begin{picture}(0,0)
\put(-10,458){$\dot{\Phi}_{2}$}
\put(234,349){$\Phi_{2}$}
\put(38,129){$\dot{a}$}
\put(206,-14){$a$}
\end{picture}
\newpage

\
\begin{figure}[h]
\vspace{70mm}
\includegraphics{mixprl3.out.ps}
\vspace{30mm}
\caption{ A detail of Fig. 13. where the dimension is $d_{B}=1.484\pm 0.005$.}
\end{figure} 

\vspace*{-30mm}
\begin{picture}(0,0)
\put(38,127){$\dot{a}$}
\put(206,-25){$a$}
\end{picture}

\end{document}